\documentclass[prb,twocolumn,showpacs,superscriptaddress]{revtex4-1}        
\usepackage{epsfig}
\usepackage{dcolumn}
\usepackage{epstopdf}
\usepackage{graphicx}
\usepackage{color}

\newcommand{\wn}{cm$^{-1}$}
\begin{document}

\title{Multi-gap absorption in CaCu$_{3}$Ti$_{4}$O$_{12}$ and  the predictivity of  ab initio  methods}

\author{Francesco Ricci}\affiliation{CNR-IOM and Dipartimento di Fisica, Universit\`a di Cagliari, Cittadella Universitaria, Monserrato, 09042 Cagliari, Italy}
\author{Paola Alippi}\affiliation{CNR-ISM, Istituto di Struttura della Materia, Consiglio Nazionale delle Ricerche, Via Salaria km 29.5 CP 10, 00016 Monterotondo Stazione, Italy}
\author{Alessio Filippetti}
\author{Vincenzo Fiorentini}\affiliation{CNR-IOM and Dipartimento di Fisica, Universit\`a di Cagliari, Cittadella Universitaria, Monserrato, 09042 Cagliari, Italy} \date{\today}
\begin{abstract} 
We report the electronic properties of the quadruple perovskite CaCu$_{3}$Ti$_{4}$O$_{12}$ as obtained via several density-functional based methods, and propose a  new interpretation of optical experiments  to the effect that four distinct transitions  (centered around 0.7, 1.5, 2.5, and 3.5 eV) contribute to the spectrum. The comparison with experiment is satisfactory, especially after we account for the effects of spin disorder, which does not close the fundamental gap but suppresses the transition intensity. We find that  some of the methods we employ tend to overestimate considerably the gaps for standard values of the respective adjustable parameters.
\end{abstract}
\maketitle
\noindent
\section{introduction}

 The popular line ``If it's  been  measured, why are you calculating it ?'' attributed to Volker Heine  emphasizes the need for electronic structure theory  to harness its predictive and interpretive potential. In this paper, we revisit the low-energy optical properties of the quadruple perovskite CaCu$_{2}$Ti$_{4}$O$_{12}$ (CCTO henceforth,  risen to popularity a decade ago\cite{ccto} because of its anomalous dielectric response), heeding the  advice implicit in Heine's remark in two distinct respects. Firstly,  theoretical predictions predated reliable experiments, and here we  provide a new and improved interpretation  of the latter.
Secondly, electronic structure theory often revisits known results to provide additional insight and to validate its  predictive power in retrospect. In this spirit we apply to CCTO several {density-functional-theory (DFT)} state-of-the-art methods, which yield a mixed bag of good and bad   news.   Some advanced methods appear to be struggling, while  others yield satisfactory agreement with experiment.

\subsection{Motivation}
Early experimental reports\cite{ccto} on CCTO circa 2002 had suggested a fundamental gap  in excess of 2.5 eV. Values as low as 0.2 eV obtained in {DFT} local-density-approximation (LDA) calculations\cite{vanderbilt} were, not unreasonably, discounted in view of the known gap underestimation problem of local and semilocal functionals. Looking at the LDA bands, however, we realized that the lowest gap might be a low-energy transition between localized and  predominantly Cu-like states, rather than the natural dipole-allowed  transition between O $p$ valence and Ti $d$ conduction bands, a situation analogous to other Mott-like insulators.\cite{altrimott} Therefore, in 2006 we used\cite{noi2} self-interaction corrected LDA (PSIC),\cite{fs} known by then  to reproduce  quite accurately the gaps in many Mott-like cuprates,\cite{ff} to find out how beyond-LDA bands would look like in the Cu-dominated gap region.  We found that the fundamental gap (indirect, forbidden, and between mostly Cu-like  states) was only about 0.6 eV, moderately increased in absolute value over the LDA value. This surprising result seemed to point to smaller-than-usual correlation effects in the nearly filled 3$d$ Cu(II) shell; put differently, the on-site interaction, which self-interaction corrections largely restore to its correct size, appeared to be rather more screened at  Cu sites in CCTO than in other magnetic Cu oxides.\cite{ff,ybcochain}

Systematic experiments (see Sec.\ref{compexp} below) first appeared in 2008, when Kant {\it et al.}  inferred\cite{lunk,lunk2} from optical conductivity an electronic structure qualitatively matching  that suggested  by LDA\cite{vanderbilt} and, to a somewhat larger extent, by PSIC,\cite{noi2} with weak Cu-dominated transitions starting at about 0.5-0.7 eV. In 2011,  a different picture was proposed,\cite{jacs} to the effect that the ``Cu-Cu'' transition would start at about 1.8 eV,  based on reflectivity measurements   interpreted via GGA+U (Generalized Gradient Approximation +U). The calculations used a U--J parameter  much smaller than the usual value for Cu oxides (reminding us of the low-correlation argument), yet it produced a fundamental gap much larger than previously obtained by PSIC. This suggested that  it would be a good idea to revisit and expand our previous  investigation applying further advanced methods to CaCu$_{3}$Ti$_{4}$O$_{12}$ to help sort out the matter and provide a  robust interpretation.

In this work, we discuss the  electronic properties of  CCTO based on results from several  different DFT-based methods. We eventually  propose an interpretation of experiments, as well as  further experimental tests, based on one of them, the variational  PSIC method\cite{vpsic} (VPSIC henceforth). Our conclusion in summary is that CCTO has a multifold interband absorption due to its unusual Cu-induced upper-valence and  lower-conduction band structure, and that the fundamental transition peaks around 0.7-0.8  eV (1500 nm), while the absorption peaking  at 1.8 eV (700 nm) is an O $p$ valence band to Cu $d$ conduction band transition, at variance with the previous interpretation.\cite{jacs} More intense absorptions between 2.5 to 4 eV  are due to transitions into the higher-lying Ti $d$ conduction band.  The fundamental gap is a Mott-like gap in the sense that the system has an odd electron count,  the gap open between strongly localized and spin-polarized states, and would not exist in the absence of magnetic moments. We account for spin disordering in the paramagnetic (PM) phase in which most measurements are performed: the fundamental gap survives unscathed  the breakdown of magnetic order, but  the intensity of the transition across that gap is suppressed. As a test of the suggested picture, we  point out features that should be observable in low- vs high-temperature optical and energy-loss spectroscopy experiments.

Besides VPSIC and GGA/LDA, we  calculate the electronic structure using GGA+U,   hybrid functionals, and many-body perturbation theory.
 Our theory-experiment comparison indicates that hybrid and GGA+U  end up quite far from experiment, overestimating severely the gaps, whenever standard values are used for  the adjustable parameters they depend on, a conclusion that has obvious methodological implications. Also, the corrections to the eigenvalues of local or semilocal functionals  provided by VPSIC are close to those of  non-self-consistent one-shot GW, suggesting that the "beyond-local" correlation is described by VPSIC with  similar accuracy.

\section{Methods}
\label{methods}

As usual in the business of ab initio optical properties, we elect to interpret the eigenvalues and eigenvectors of Kohn-Sham equations as quasiparticle energies and states. This is  justified by the  Kohn-Sham equations being formally identical to Hedin-Lundqvist quasiparticle equations\cite{hedin} if the self-energy $\Sigma$ is identified with the exchange-correlation potential;  for LDA, e.g., $\Sigma_{\rm LDA}$({\bf r},{\bf r$'$},$E$)$\equiv$$\delta$({\bf r}--{\bf r$'$})$\delta$($E$) V$_{\rm LDA}^{xc}$({\bf r}),  and similarly for functionals containing some degree of  non-locality and implicit energy dependence such as hybrids or PSIC.\cite{ff} 
It is obviously interesting, therefore, to compare results obtained by different exchange-correlation functionals. This is done in Sect.\ref{ggauhse}, in particular \ref{altri2}; In the same spirit, we also discuss in Sect.\ref{gws} ``many-body'' corrections to semilocal functionals, both empirical and based on  G$_{0}$W$_{0}$  non-self-consistent many-body perturbation theory.\cite{gw} 

\begin{figure}[h]
\center\includegraphics[width=7 cm]{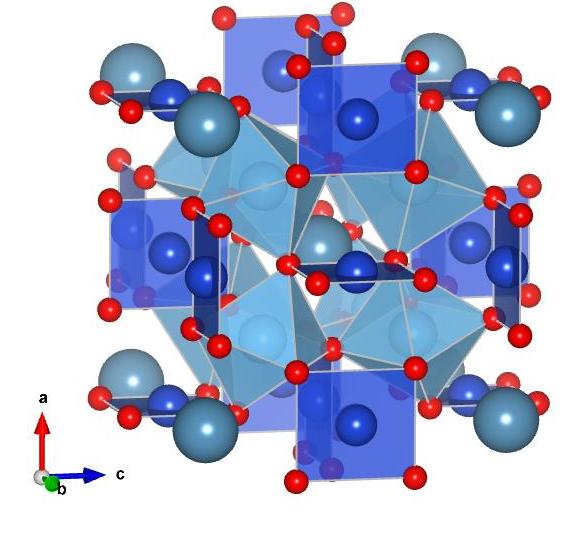}
\caption{(Color online)   Crystal structure of CCTO.}
\label{figstr}
\end{figure}

In recent years,  methods going beyond the local or semi-local approximation   have   become more affordable, and we are in the position to evaluate their relative merits under the assumption stated above. The optical conductivity, extinction coefficient, and  electron-energy-loss function are extracted from the dielectric function $\tilde{\varepsilon}$($\omega$) calculated within the random phase approximation from the joint density of states obtained with the variational version of pseudo self-interaction-corrected LDA (VPSIC).\cite{fs,ff,vpsic} We also calculate gaps and transitions with
  Ceperley-Alder LDA,\cite{ca} Perdew-Becke-Ernzerhof GGA,\cite{pbe} the Dudarev {version of the} GGA+U,\cite{dudarev}  the Heyd-Scuseria-Ernzerhof (HSE) hybrid functional.\cite{hse} To avoid bias due to changes in {volume}, we use the cubic magnetic primitive cell (40 atoms) at the experimental lattice constant of 7.38 \AA\, with internal coordinates optimized with GGA, {imposing a threshold of 0.01 eV/\AA\, on force components. Since CCTO is cubic, the positions of Ti, Cu and Ca are fixed by symmetry;  Ti-O octahedra and Cu-O plaquettes are "rigid" and all identical geometrically (see Fig.\ref{figstr}), so the O positions are determined by just the Ti-O and Cu-O distances (1.959 \AA\, and  1.963 \AA, respectively).}

VPSIC uses ultrasoft pseudopotentials and plane waves in a home-made custom code  with cutoff 475 eV. All other methods are implemented in VASP\cite{vasp} and use the PAW\cite{paw} method with cutoff 400 eV.  We employ standard  k-point meshes (4$\times$4$\times$4 for self-consistency and up to 12$\times$12$\times$12 for density-of-states or dielectric function calculations). {In the optics calculations, the imaginary part $\epsilon_2$ of the dielectric function is calculated directly, whereas the  real part  is obtained via the Kronig-Kramers relations (for standard relations, see e.g. 
Ref.\onlinecite{grpa}); we use up to 2000 bands in the summation over empty states (for the G$_0$W$_0$ calculations, see the discussion in Sec.\ref{gws}), which are amply sufficient to converge  both the imaginary and real parts of the dielectric constant at the energies of interest (below about 10 eV).}

Some of the techniques employed  involve adjustable parameters. The Dudarev GGA+U version depends on parameter U--J, applied to  Cu $d$ states. The HSE hybrid, in turn, is tuned via the fraction $\alpha$ of screened Hartree-Fock exchange  and the screening cut-off wavevector $\mu$. In the current VPSIC  formulation,  screening of the self-interaction  by the environment is quantified by a constant which may be treated as a  parameter. However, we keep it fixed, as in all past applications, at a value based on a Slater-transition-state concept explained in Ref.\onlinecite{fs}.

\begin{figure}[h]
\center\includegraphics[width=8 cm]{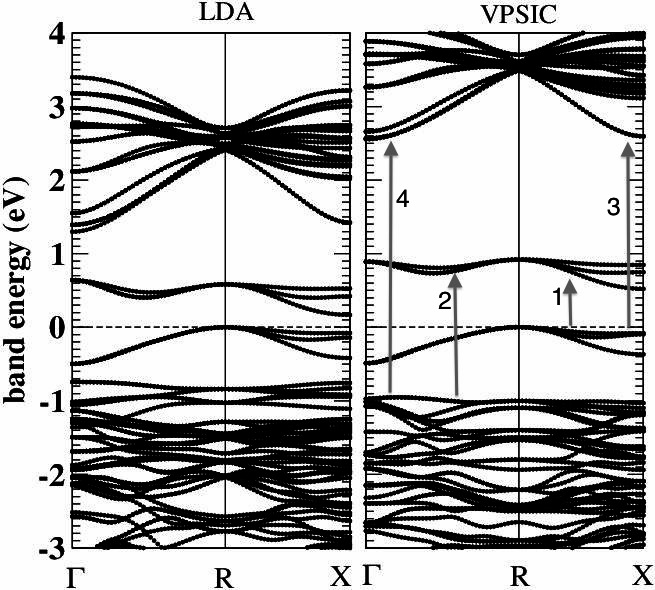}
\caption{LDA vs VPSIC bands of CCTO. Arrows indicate schematically the four transitions discussed in the text.}
\label{fig1}
\end{figure}

\section{VPSIC results  vs. experiment}

This Section  compares the VPSIC optical functions of CCTO with two distinct sets of experimental data. In Sec.\ref{ggauhse} we will examine and discuss the gaps  provided by the other methods. We will be switching  units several times to ease the  comparison with  experiment.

\subsection{Band structure} 

\begin{figure}[h]
\center\includegraphics[width=8 cm]{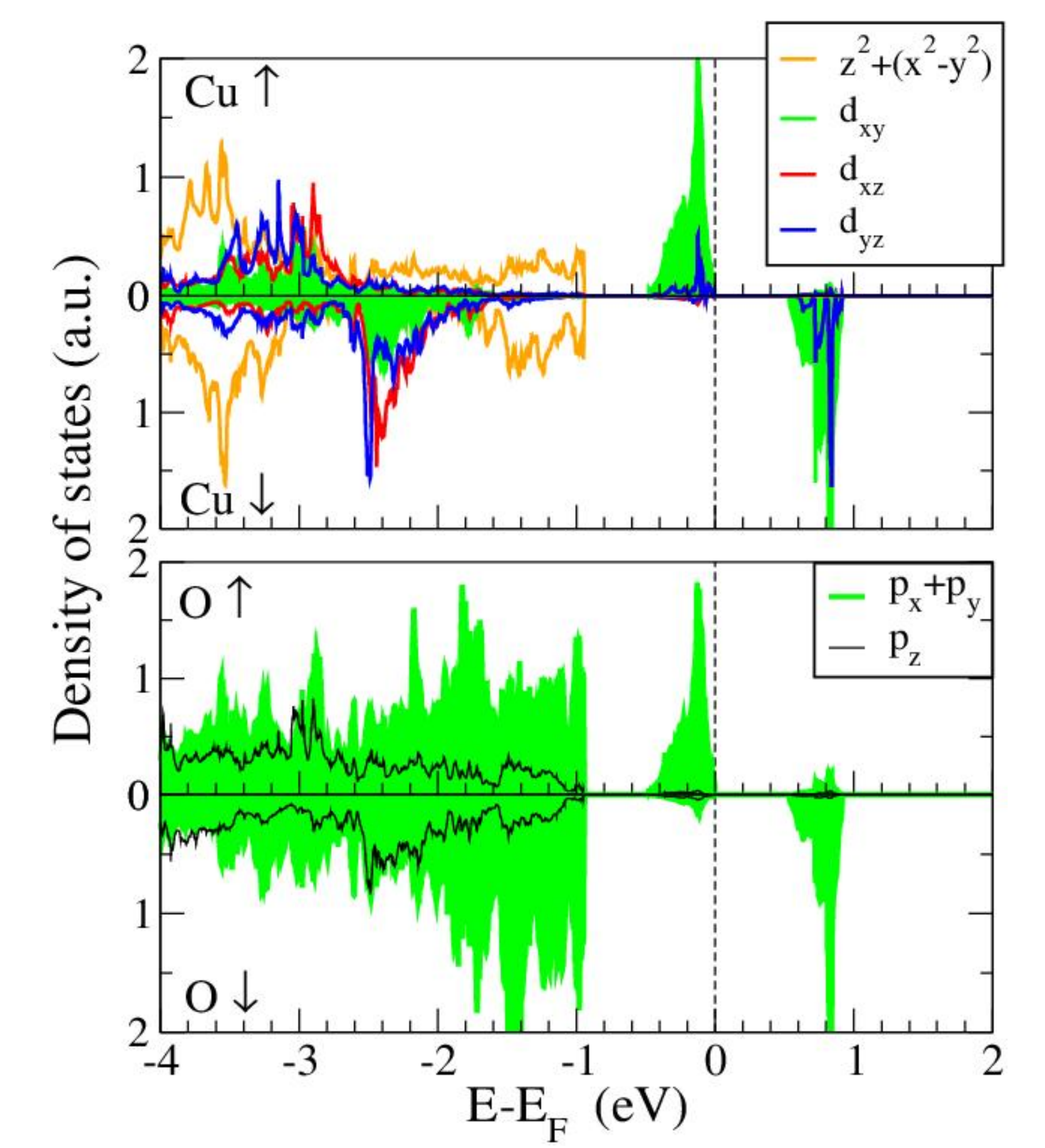}
\caption{(Color online)  {DOS of CCTO  from VPSIC in the near-gap region, projected on orbitals of the Cu and O atoms of a plaquette in the $x$$y$ plane (see Fig.\protect\ref{figstr}).}}
\label{fig2}
\end{figure}

\begin{figure}[h]
\center\includegraphics[width=8 cm]{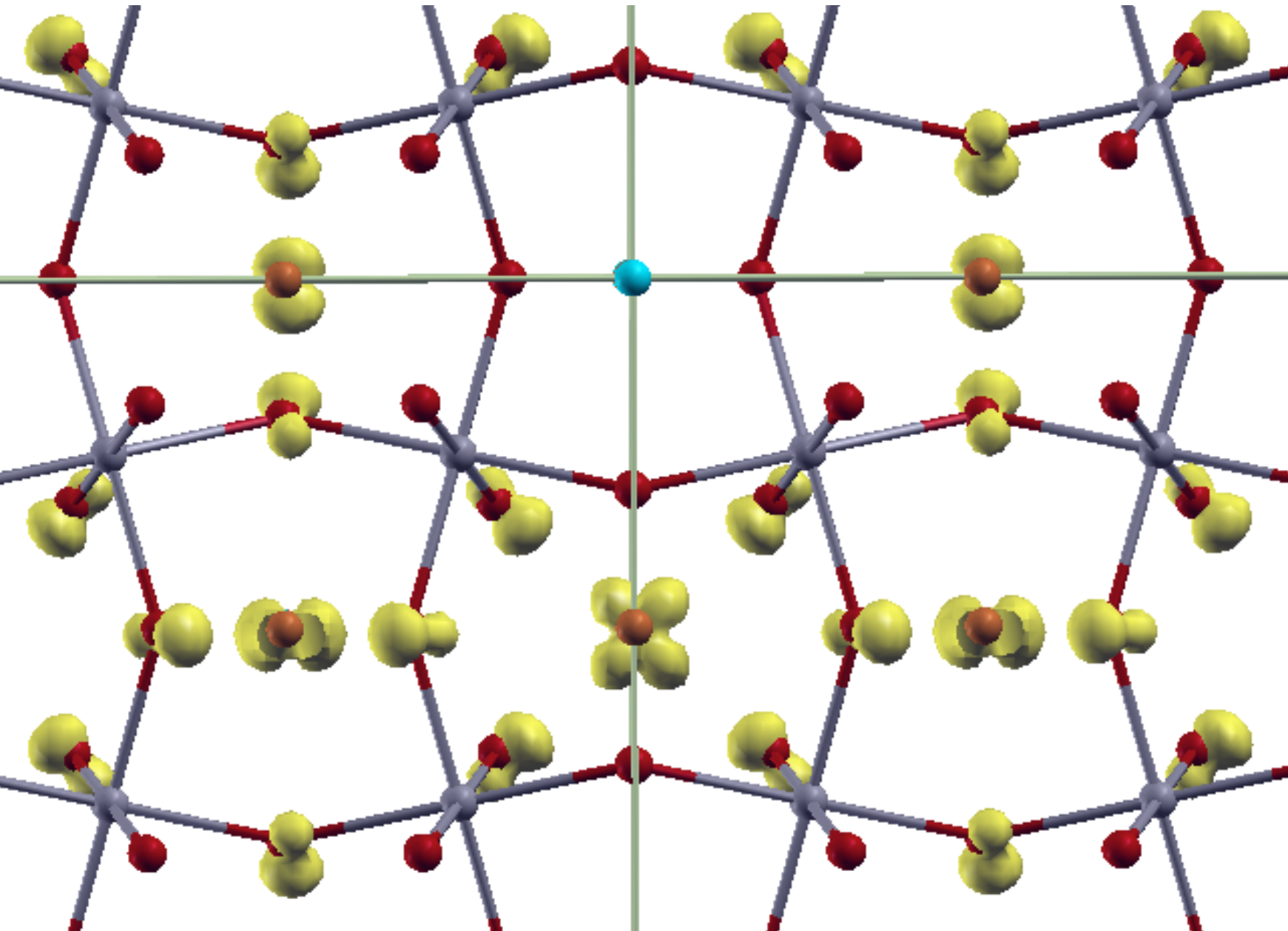}
\caption{(Color online)  {Magnetization density in CCTO seen along the $z$ axis. All three kinds of Cu-O plaquettes are visible. The Cu onto which  the DOS in Fig.\ref{fig2} is projected is the one at bottom center, with the $x$$y$-orbital--shaped density.}}
\label{figmag}
\end{figure}

CCTO is G-type antiferromagnetic (AF) on the Cu lattice with a Ne\'el temperature of 25 K.
Its band structure  obtained by LDA and VPSIC is displayed in Fig.\ref{fig1}. The LDA and VPSIC bands are rather similar in structure and energy separation. The near-gap bands connected by the transition marked `1' in Fig.\ref{fig1} are dominated by O-hybridized Cu-like spin-polarized states. The top valence band and the bottom conduction band are fully spin-polarized, and as can be seen in
the  orbital- and site-projected density of states (DOS)  as obtained by VPSIC in Fig.\ref{fig2}, their projections on any given Cu site have opposite  polarization. As usual, the removal of self-interaction tend to increase all the gaps. The largest increases are found for  the transitions marked `3' and `4' in Fig.\ref{fig1} to  the upper conduction band of predominantly Ti character, i.e.   for the standard charge transfer gaps. 
The local orbital character of the near-gap spin-polarized states is completely determined by Cu $d$ and ligand O's in each plaquette (see Fig.\ref{figmag}), in accordance with the DOS of \ref{fig2}.

As far as optical absorption is concerned, the band structure in 
  Fig.\ref{fig1}  suggests that four distinct relevant absorptions are expected,  marked `1' to `4'. The first transition is between O-hybridized Cu-like bands, between 0.5 and 0.9 eV (2000-1000 nm, 4000-7000 cm$^{-1}$). A large joint DOS is expected due to  extended parallel-band sectors  especially around the X point; also, despite the similar character of the initial and final states, the matrix elements should not be suppressed, because  this is an intersite transition (intrasite transitions are forbidden by the spin conservation selection rule). The second transition is
between valence O $p$ and low-conduction Cu $d$ bands in the range  
1.4-1.9 eV (800-650 nm, 11000-15000 cm$^{-1}$), which is not expected to be suppressed selection-rule-wise. 
The third absorption is Cu $d$ upper-valence to Ti $d$-O $p$ conduction at 2.6-3.0 eV (500-400 nm, 21000-24000 cm$^{-1}$), which is expected of average intensity; finally, an intense O $p$-Ti $d$ transition should start at 3.4 eV (350 nm; 27500 cm$^{-1}$).

The lower conduction band of Cu character is  affected only weakly by self-interaction corrections, and accordingly the lower-energy O $p$--Cu $d$ and Cu $d$--Cu $d$ transitions `1' and `2' change moderately compared to LDA. This feature is key to our interpretation, and, as we will see in Sec.\ref{ggauhse}, it is not shared by other methods.
The small magnetic fundamental gap `1' may be labeled  as Mott-like, since it depends on the interplay of spin polarization, Hund coupling, and on-site repulsion, and it is coherent with the textbook definition  $U$--c$t$, i.e. it includes (thanks to self-interaction removal) the cost U of adding an electron in the empty $d$ state as well as the hopping $t$, which is included in the band width.\cite{nota} Since even LDA finds this gap (albeit barely), the on-site correlation acting on these states must not be especially strong.
Further, the stronger O $p$-Cu $d$ hybridization plays a role in reducing the gap  in CCTO compared to  e.g. in  YBa$_2$Cu$_3$O$_{6}$, whose gap\cite{ybcochain}  is 1.2 eV.

\subsection{Paramagnetic vs antiferromagnetic phase}
\label{PMvsAF}

Before delving into the comparison, we need to discuss the role of spin disorder. The experiments  we consider are done well   above the   N\'eel temperature T$_{\rm N}$=25 K of the AF phase of CCTO, and therefore  probe the paramagnetic (PM) phase. Since PM CCTO  is insulating experimentally, it is most likely a collection of thermally disordered, randomly oriented Cu moments, rather than a zero-moment Pauli-type PM (in the latter phase, CCTO is found to be a metal). The fundamental transition, labeled `1' in Figs.\ref{fig1} and below, is  between Cu-like spin-polarized states (see  Fig.\ref{fig1} and Fig.\ref{fig2}) and is spin-selective in the sense that
  spin-allowed transitions only occur with matrix elements involving same-spin sites of the Cu lattice. In the PM, we expect the intensity  of  transition `1' to  be reduced, because spin mixtures will be involved. To expect a good match with experiment, this suppression should be assessed and accounted for.

{We do not aim at sampling in detail the PM configurations (which is outside our scope and well beyond ``naive'' sampling techniques); rather we need to show a) that a gap survives in the misaligned-spin, i.e. non-AF-ordered, system when moments are non-zero, and b) that the intensity of the transition between the spin-polarized states is suppressed. Point a) is worth making directly; the notion that  ab initio methods can obtain good gaps and magnetism in correlated systems where LDA or GGA fail (as the PSIC does, e.g. for YBCO,\cite{ybcochain} CuO,\cite{cuo,cuo2} LaTiO$_3$,\cite{vpsic} LaNiO$_3$/LaAlO$_3$ superlattices\cite{danilo} etc.) is generally considered with suspicion because  of the almost ubiquitous assumption of magnetic order in such systems. Our simplified  disordered-moments PM, in fact, turns out to have a gap; a similar conclusion was drawn earlier for MnO\cite{luders} based on essentially the same electronic-structure technique (and a much better spin-disorder sampling).  As to point b), in the PM  the spin states are mixtures, i.e. spin projections are no longer just unity or zero, referring to a given quantization axis. Intersite transitions, that were spin-conserving in the AF, will thus be suppressed in the PM, whereas on-site transitions between formerly opposite-spin states will gain non-zero amplitude. While the latter are expected to be suppressed by the dipole selection rule, it is appropriate to explore how these two effects play out quantitatively.}

To assess the degree of intensity suppression of   absorption `1' in the PM phase, we perform non-collinear-magnetization calculations whereby  the six Cu moments in the primitive cell  are  oriented randomly, but directionally constrained to give a total magnetic moment of zero-- that is, mimicking in effect one of the thermally accessible configurations of the disordered paramagnet.  {The non-collinear spins in the PM model are constrained by a penalty function: if that penalty  is turned off, the AF ground state is recovered. Spin-orbit coupling is not included in these calculations. 
Since the low energy bands obtained with semi-local-functionals  and self-interaction corrections  are quite similar (see Fig.\ref{fig1}), we use GGA to access the non-collinear and penalty-function features of VASP. As we are only interested in the effect on the lowest gap, we display the low energy portion of the GGA absorption for the PM (aligned to match the VPSIC gap) together with the VPSIC absorption.}

\begin{figure}[h]
\center\includegraphics[width=7 cm]{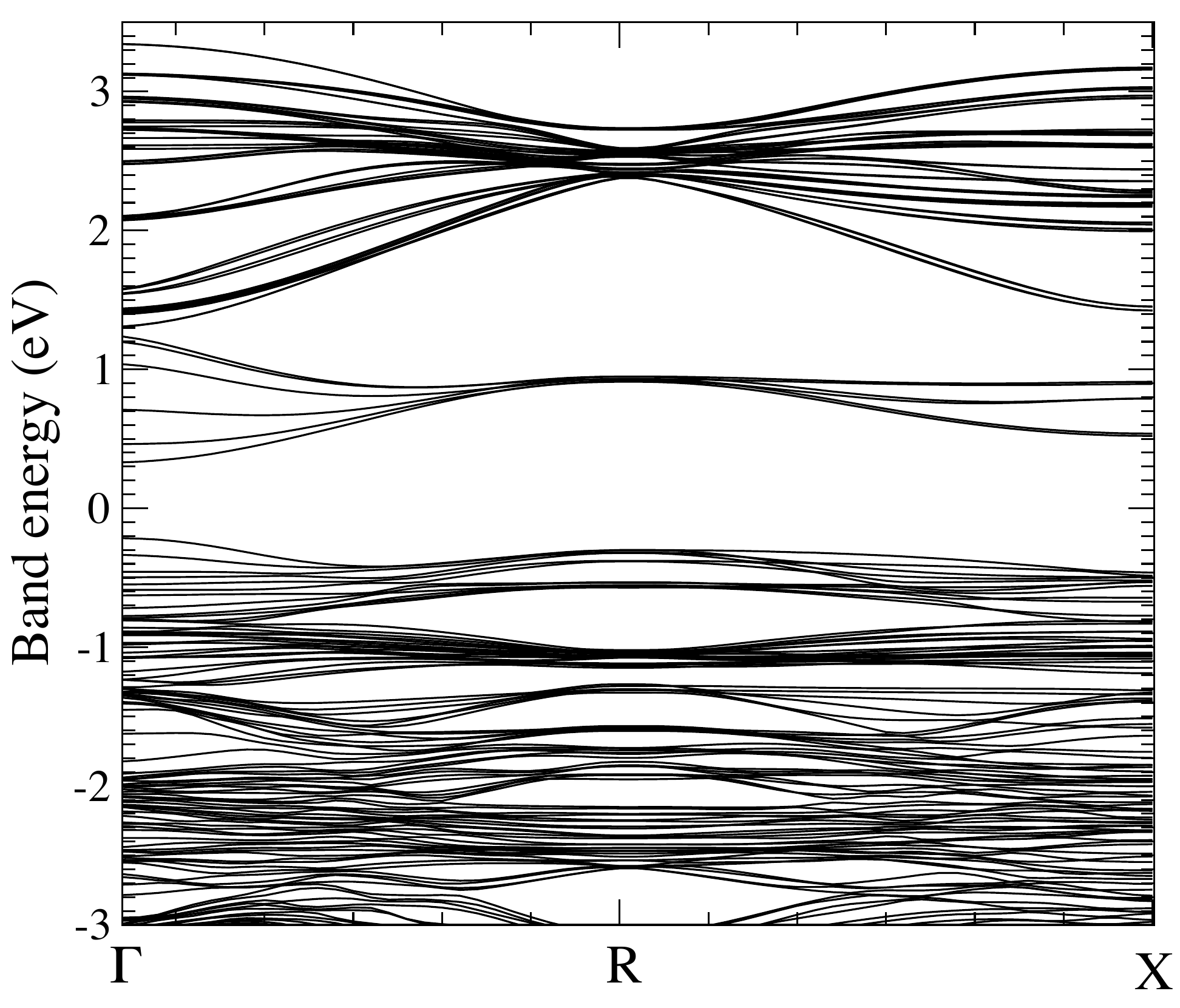}
\caption{GGA bands for the CCTO cell with non-collinear spins mimicking the disordered PM.}
\label{bands-PM}
\end{figure}

 The key result, as can be seen from the bands in Fig.\ref{bands-PM}, is that the gap remains  non-zero in the disordered PM, and close to the AF value. This should help dispel the myth that the distruction of magnetic order will lead to metallicity in ab initio calculations. Indeed,  it does not, as long as magnetic moments survive.\cite{luders} A related result relevant to our interpretation below  is that,  as we surmised,  the  fundamental absorption is indeed  suppressed in the PM compared to the AF, whereas the rest of the spectrum is practically unchanged. This improves agreement with experiments done at high temperature, as we discuss in the next Section.
We report only the low energy portion of the PM optical constants (up to about 1.5 eV), since spin disorder only affects the spin-polarized transition `1'. 
{We note in passing that the metallic Pauli-PM phase (not shown) shows typically metallic optical features such as a Drude peak at low-frequency, which are not observed in any of the experiments.}

\subsection{Comparison with experiments}
\label{compexp}

Based on  diffuse reflectance measurements, Ref.\onlinecite{jacs} suggests as lowest-energy transition an indirect-gap  absorption peaking at 700 nm, and attributes it to  transitions from the  mainly Cu-like upper valence states to the mainly Cu-like first conduction band, i.e. to transition `1' of our band structure in Figs.\ref{fig1} and \ref{fig2}. Higher-energy  intense absorptions are attributed  to O $p$-Ti $d$ dipole transitions. \cite{jacs} This interpretation is based on GGA+U calculations with U--J=6.5 eV, a rather low value for Cu oxides, which nevertheless pushes  transition `1'  up to the needed 1.7-1.8 eV (see the discussion in 
Sec.\ref{ggauhse}).

\begin{figure}[h]
\center\includegraphics[width=7 cm]{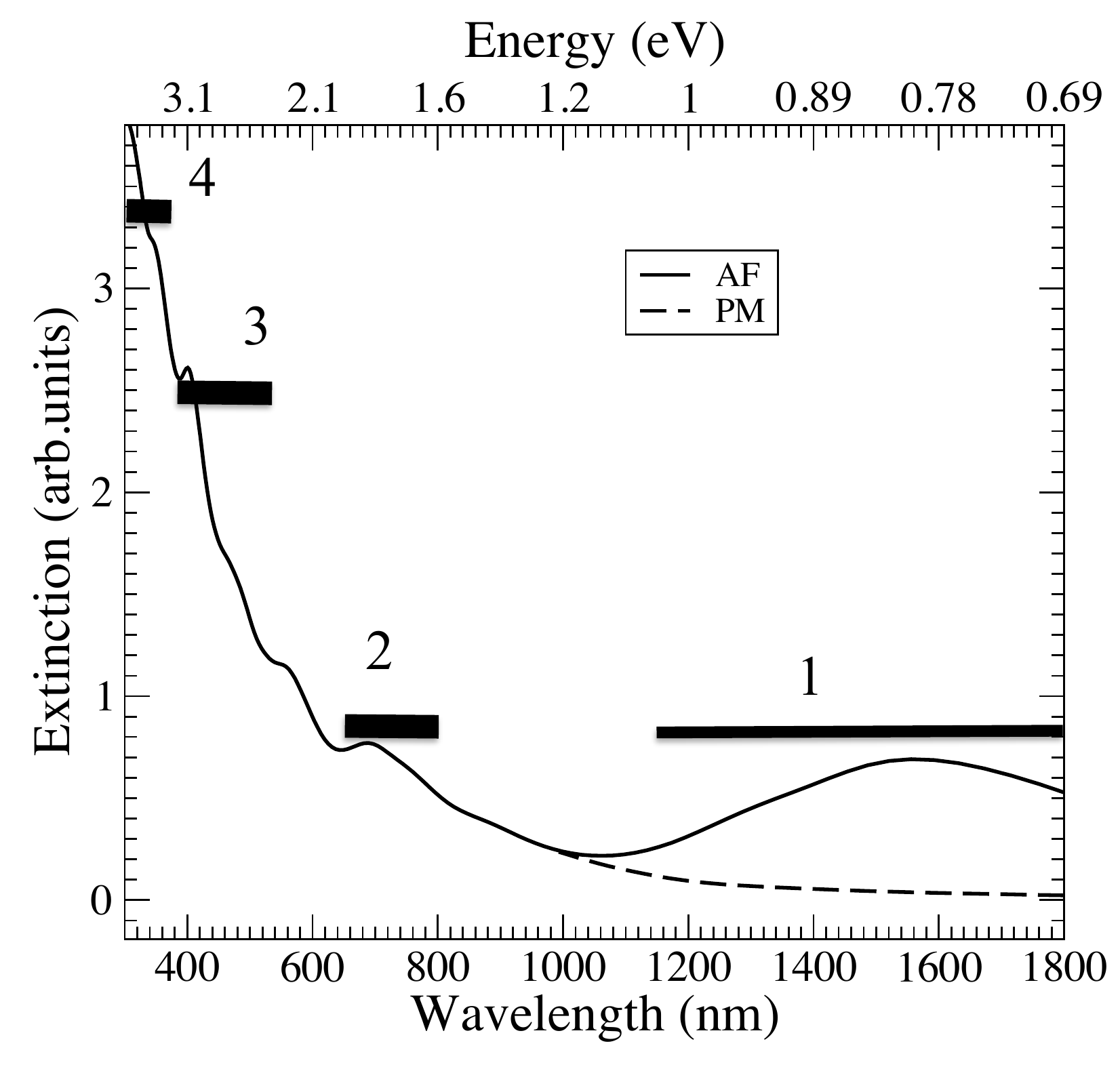}
\caption{Extinction coefficient for PM and AF CCTO calculated with VPSIC. Compare with Fig.7b (inset) of Ref.\onlinecite{jacs}.}
\label{fig3}
\end{figure}

\begin{figure}[h]
\center\includegraphics[width=7 cm]{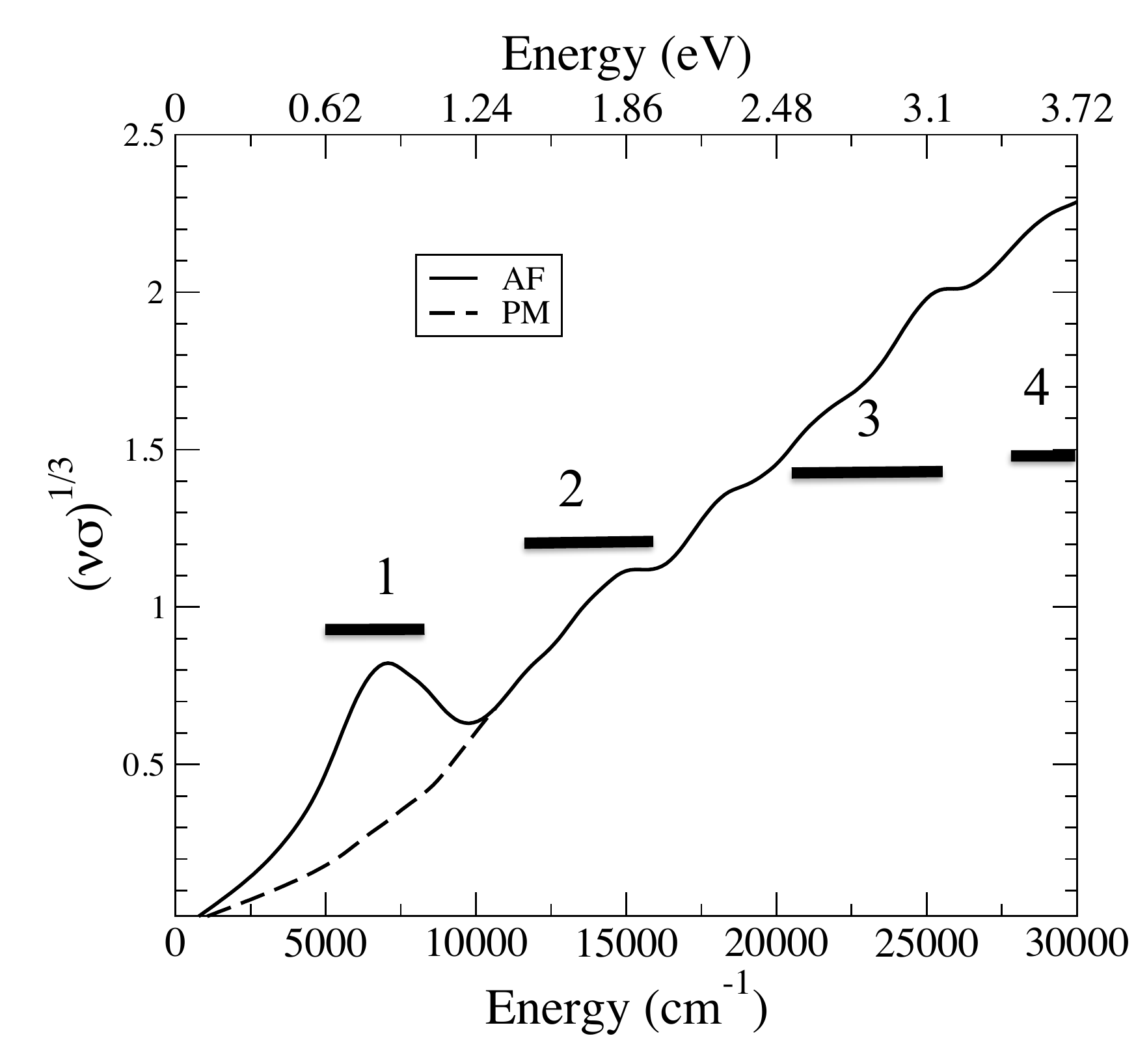}
\caption{Tauc relation for  CCTO (exponent for indirect forbidden transitions) calculated with VPSIC. Compare with Fig.\ref{lunkprivate}.}
\label{tauc}
\end{figure}

To compare with this experiment directly, we calculate the extinction coefficient (Fig.\ref{fig3}) and  conductivity (Fig.\ref{tauc} and \ref{sigma}; see the discussion below) from the    dielectric function $\tilde{\varepsilon}$($\omega$)=$\varepsilon_{1}$+$i$$\varepsilon_{2}$ (the strongest dependence is on the imaginary part $\varepsilon_{2}$).  Based on these results, and in particular  the extinction coefficient displayed in Fig.\ref{fig3} (to be compared e.g. with the inset of Fig.7b of Ref.\onlinecite{jacs}),  we propose a different interpretation than that just outlined: the peak at 700 nm is the O $p$-Cu $d$ labeled `2' in Fig.\ref{fig1} and \ref{fig3}, which is also an indirect transition; the shoulder at 400-450 nm is  the O $p$-Cu $d$ labeled `3'; and finally the main peak at 300 nm is due to the main interband O $p$-Ti $d$ transition, labeled `4'. 
 
The fundamental transition, which connects the upper valence and bottom conduction  Cu-like states and is labeled `1' in Figs.\ref{fig1} and \ref{fig3}, is instead at lower energy, peaking at about 1500 nm in the AF phase. 
However, in the PM the intensity of  this absorption is suppressed. As no experimental data were reported\cite{jacs} in this  wavelength region, it is probable that no significant signal was detected. Account for spin disorder resolves the potential discrepancy. Conversely, our result suggests that similar experiments in the AF phase at low temperature (and pure, untwinned, magnetically ordered crystals) should reveal this low-energy transition, providing  a direct experimental countercheck of our interpretation.

We now come to wide-range dynamic conductivity measurements,\cite{lunk} which also seem to  suggest a multigap spectrum. 
Tauc extrapolation at low energy is difficult due to low intensity and the probable indirect character of the transition, but a very weak indirect  transition starting at about 5000 \wn\,  can be inferred.\cite{lunk,lunk2} Another more intense transition follows at about 1.5-1.7 eV and finally the  intense allowed absorption  peaks at 3 eV.

\begin{figure}[h]
\center\includegraphics[width=7 cm]{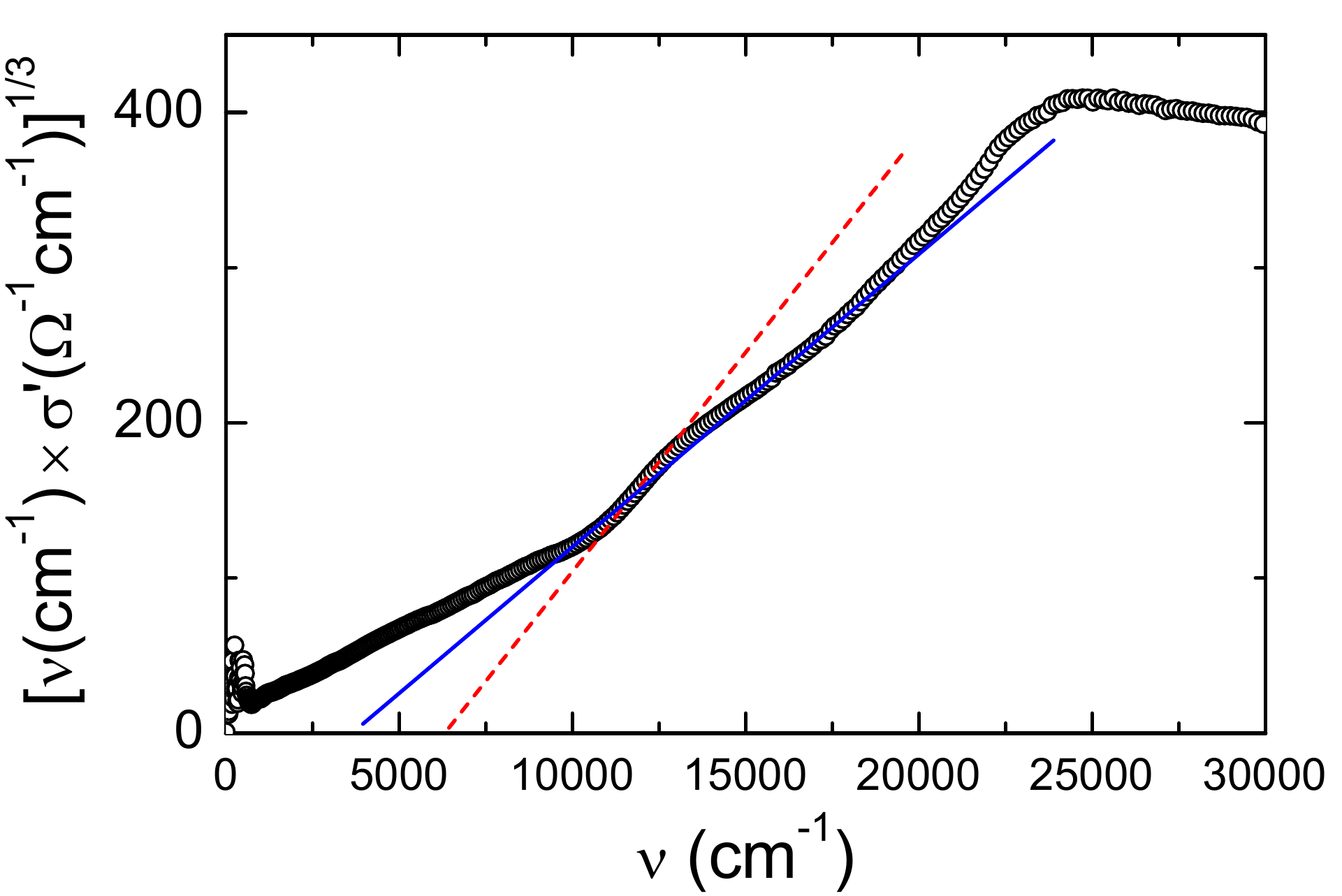}
\caption{Experimental Tauc relation\cite{lunk2} for CCTO, to be compared with Fig.\ref{tauc}. Lines are possible fits for indirect forbidden transitions whose intercepts with the frequency axis  identify the minimum gap. Figure  by courtesy of  P. Lunkenheimer.}
\label{lunkprivate}
\end{figure}

These features are reasonably well reproduced by our calculation for the PM in Fig.\ref{tauc}, as can be seen comparing with the experimental data in Fig.\ref{lunkprivate}, where the  Tauc fits suggest an onset (i.e. a minimum gap) at about 0.6-0.9 eV. This assignment is only tentative as there is no clear linear behavior over an extended frequency range.

Comparing AF and PM results, it appears that  the seemingly strongly forbidden  character of the fundamental  transition is  mainly a token of  spin disorder, rather than of interband matrix element suppression. Indeed, intrasite $d$-$d$ transitions would be suppressed by Laporte's selection rule, but here they are effectively intersite, because of the spin structure of the material; the fundamental transition in the AF is in fact quite prominent (Fig.\ref{fig3} and Fig.\ref{tauc}). 
%
Thus the same measurements  below T$_{\rm N}$  should show a marked intensity enhancement in the 5000-8000 cm$^{-1}$ (0.5-1 eV) range, providing another countercheck on our interpretation.
The same applies to the conductivity, displayed in Fig.\ref{sigma}, whose behavior for the PM compares favorably with Fig.7 of Ref.\onlinecite{lunk}.

\begin{figure}[h]
\center\includegraphics[width=8 cm]{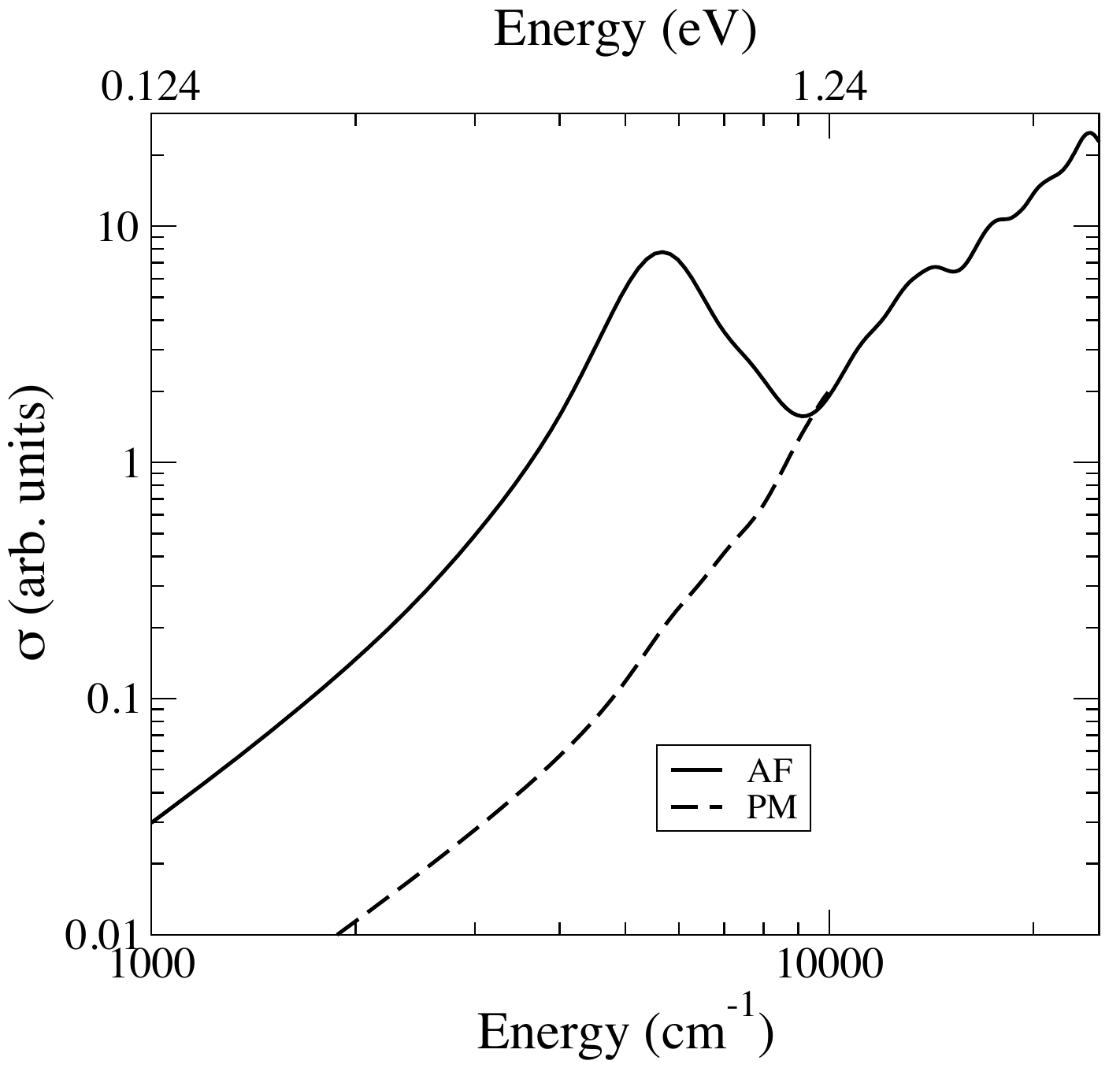}
\caption{Dynamical conductivity for CCTO calculated with VPSIC. Compare with Fig.7 of Ref.\onlinecite{lunk}.}
\label{sigma}
\end{figure}

A related result from Ref.\onlinecite{lunk} is that DC conductivity is Arrhenius-activated with a 0.2 eV characteristic energy. We attribute this simply to thermal carriers excitation across the fundamental gap. The latter is 0.6 eV in VPSIC, but we have estimated from a single GGA calculations including spin-orbit (not shown in the Figures) that the spin-orbit splitting of the upper valence and lower conduction bands (both having sizable Cu character) will reduce the gap to about 0.3 eV.

\begin{figure}[h]
\center\includegraphics[width=8 cm]{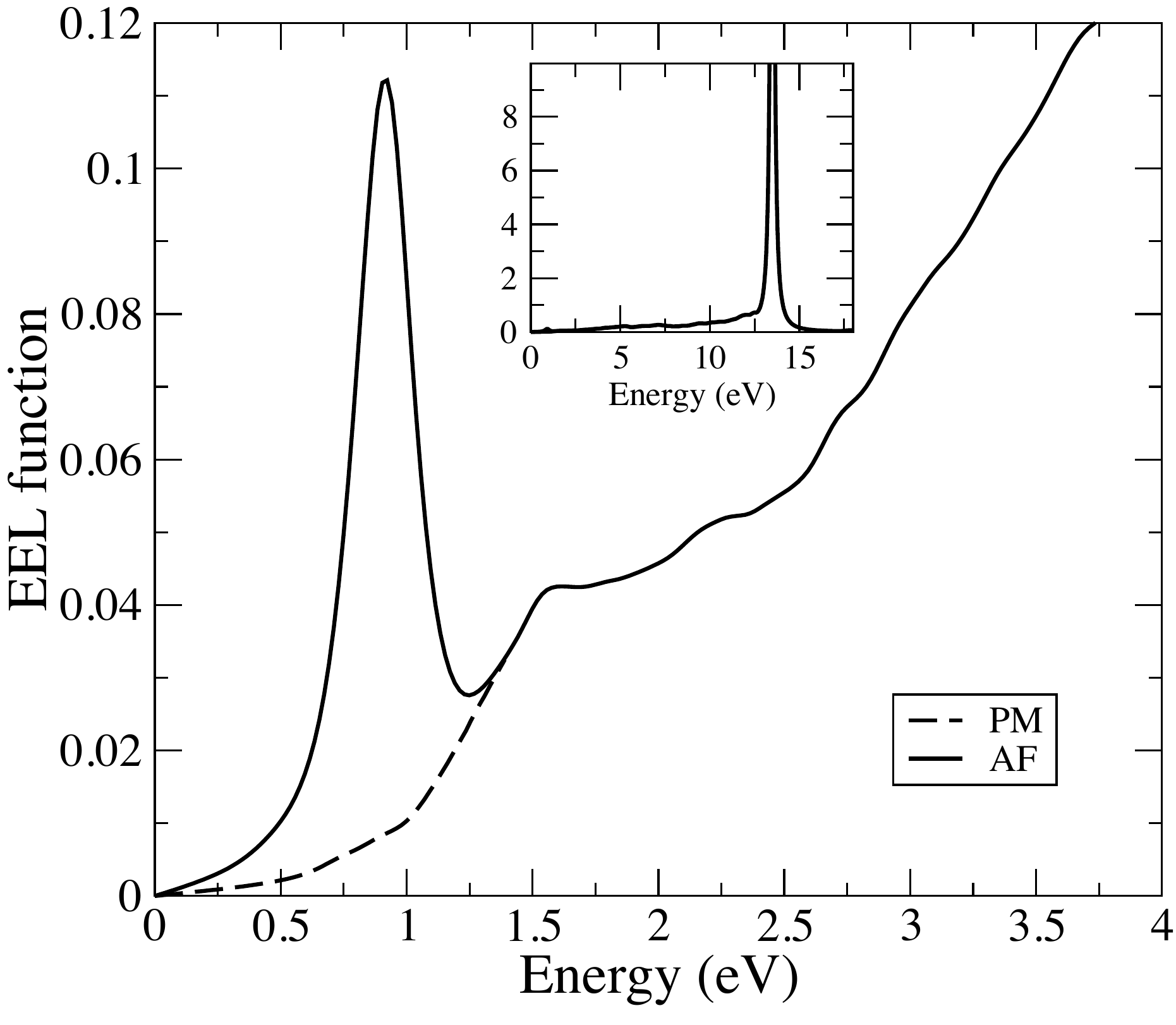}
\caption{Energy-loss function of AF and PM CCTO. The inset shows the main plasmon in the AF.}
\label{eels}
\end{figure}

In closing this Section,  we point out that beside   low-temperature optical absorption  another possible  countercheck on our suggestions is electron-energy-loss spectroscopy, again in the 0.5-1 eV  range. As shown in Fig.\ref{eels}, the energy-loss function --Im[1/$\tilde{\varepsilon}$($\omega$)] has  a marked peak at 0.9 eV in the AF (i.e. at low temperature) which is strongly suppressed in the PM (i.e. at high temperature) because of spin disorder.
The sharp main plasmon at 13.5 eV (Fig.\ref{eels}, inset) is the same in the AF and PM.

\section{Results with other functionals}
\label{ggauhse}

Having obtained a  satisfactory interpretation of the electronic structure of CCTO  using VPSIC, we  examine and compare the
 main band gaps obtained by  GGA+U and  the HSE hybrid functional.
 As mentioned earlier,  GGA+U in the version  employed here depends on the U--J parameter; we apply to the Cu $d$ shell a U--J ranging from 0 to 8 eV (a value of 8 or 9 eV is  standard\cite{ybcochain} for Cu oxides). 
 The HSE hybrid depends on the fraction $\alpha$ of screened Hartree-Fock exchange  and the  screening cut-off wavevector $\mu$: we consider $\alpha$=0, 0.1, and 0.25, the latter being {the proper HSE recipe} (while varying $\alpha$, we keep the standard $\mu$=0.2 \AA$^{-1}$); then we  explore values of $\mu$ from 0.1 to 0.5 \AA$^{-1}$, at  the standard $\alpha$=0.25 (large $\mu$ means strongly screened Fock exchange at all wavevectors, recovering GGA as $\mu$$\rightarrow$$\infty$).  
{The HSE standard value has a theoretical foundation in the formulation of the functional, and has the merit of being system-independent.  That said,  we deem this exploration worthwile, as  the $\alpha$  and $\mu$ parameters  have been, on occasion, adjusted away from their standard value to cure various different issues in cuprates and titanates.}
 
 Finally we discuss quasiparticle corrections from GGA-based G$_{0}$W$_{0}$ many-body perturbation theory; G$_0$W$_0$ has no adjustable parameter per se, but uses the GGA bands to evaluate the Green's function and screened interaction, whence the quasiparticle energies, and is not self-consistent.
 
\subsection{Parameter dependence of main gaps in HSE and GGA+U}
\label{altri2}

\begin{figure}[h]
\center\includegraphics[width=7 cm]{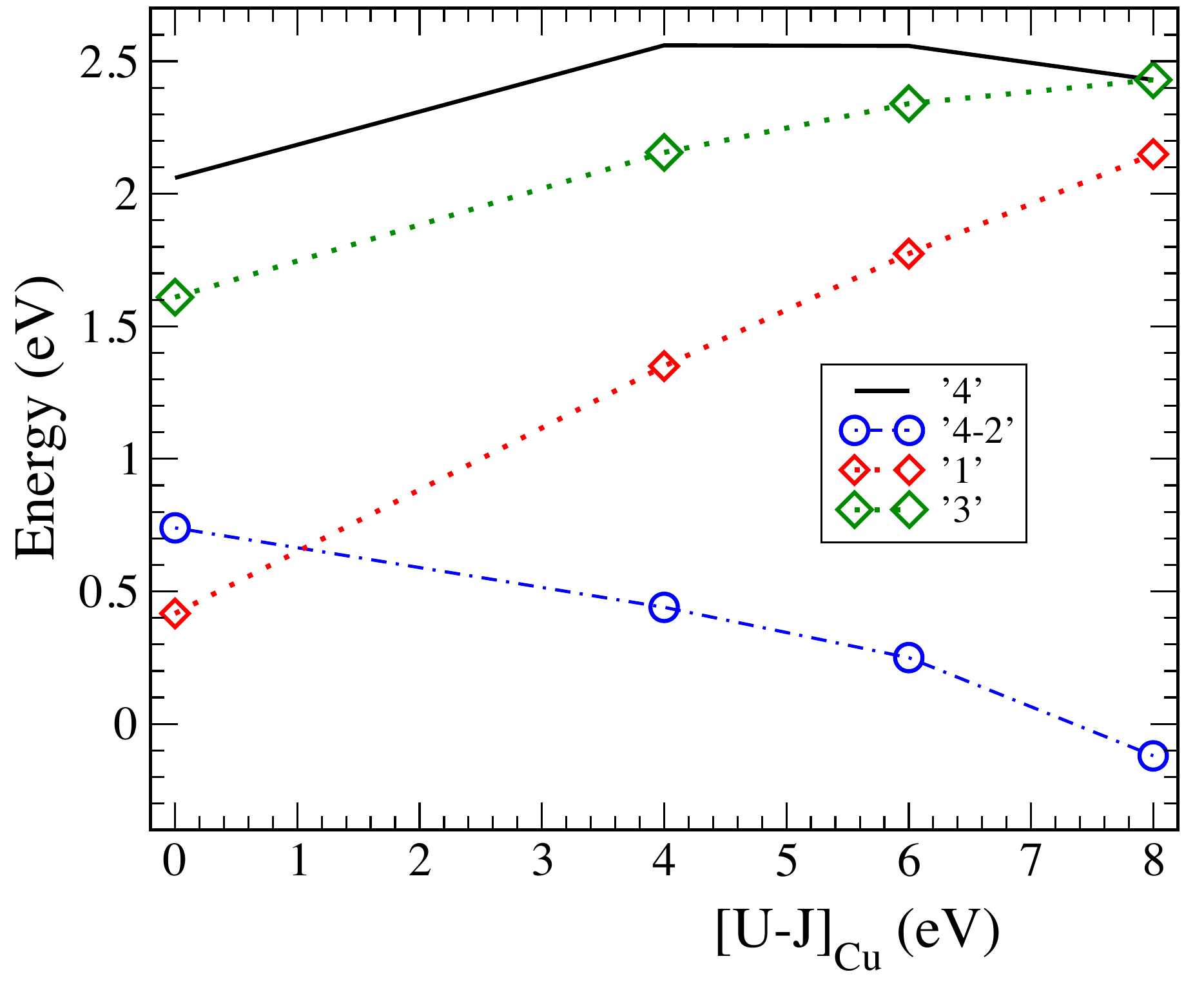}
\caption{(Color online) GGA+U transitions (see text, and the scheme in Fig.\ref{fig1})  vs U--J for Cu.}
\label{fig5}
\end{figure}

\begin{figure}[h]
\center\includegraphics[width=7 cm]{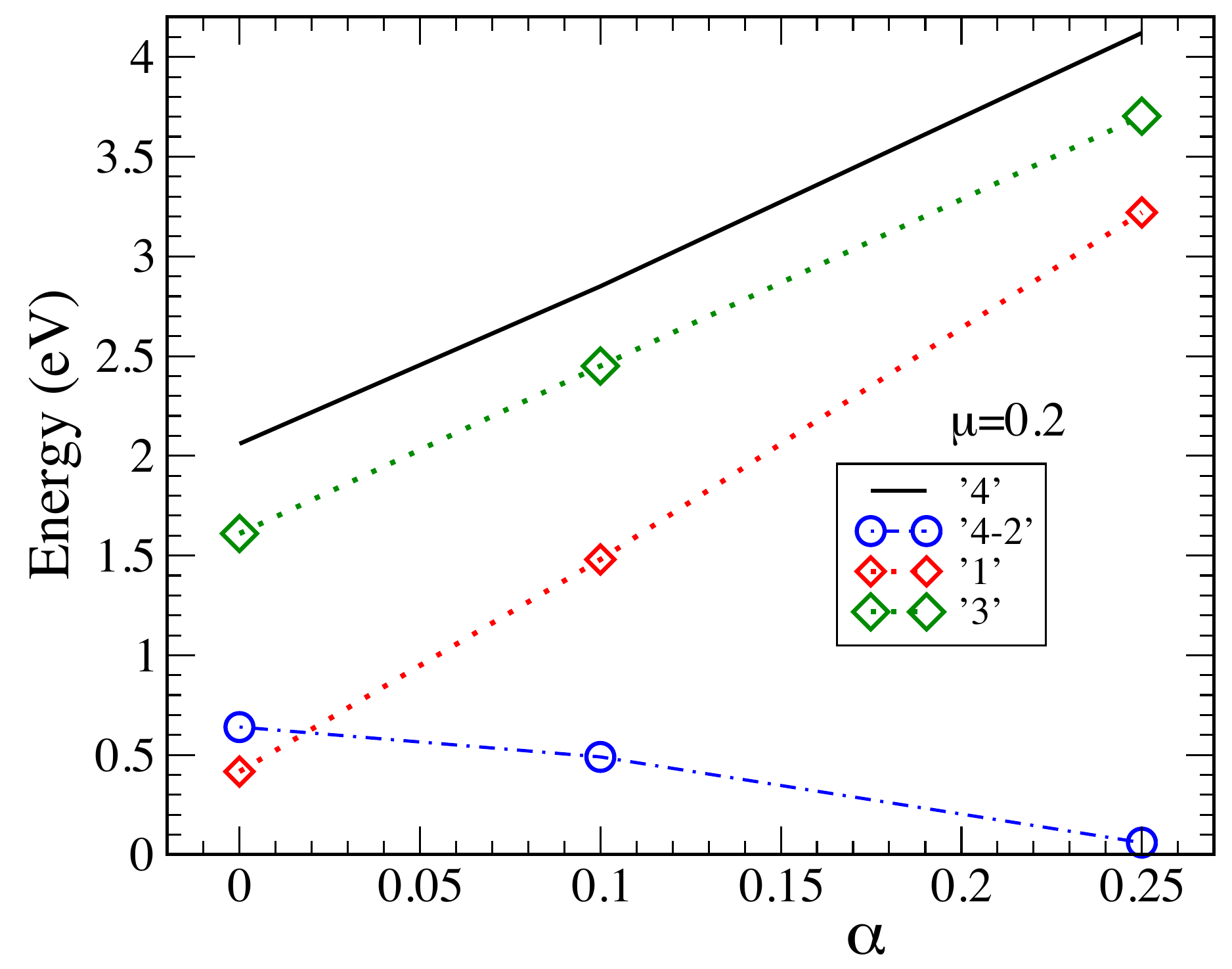}
\caption{(Color online) HSE transitions (see text, and the scheme in Fig.\ref{fig1}) vs $\alpha$, with $\mu$ fixed at 0.2 \AA$^{-1}$.}
\label{fig6}
\end{figure}

\begin{figure}[h]
\center\includegraphics[width=7 cm]{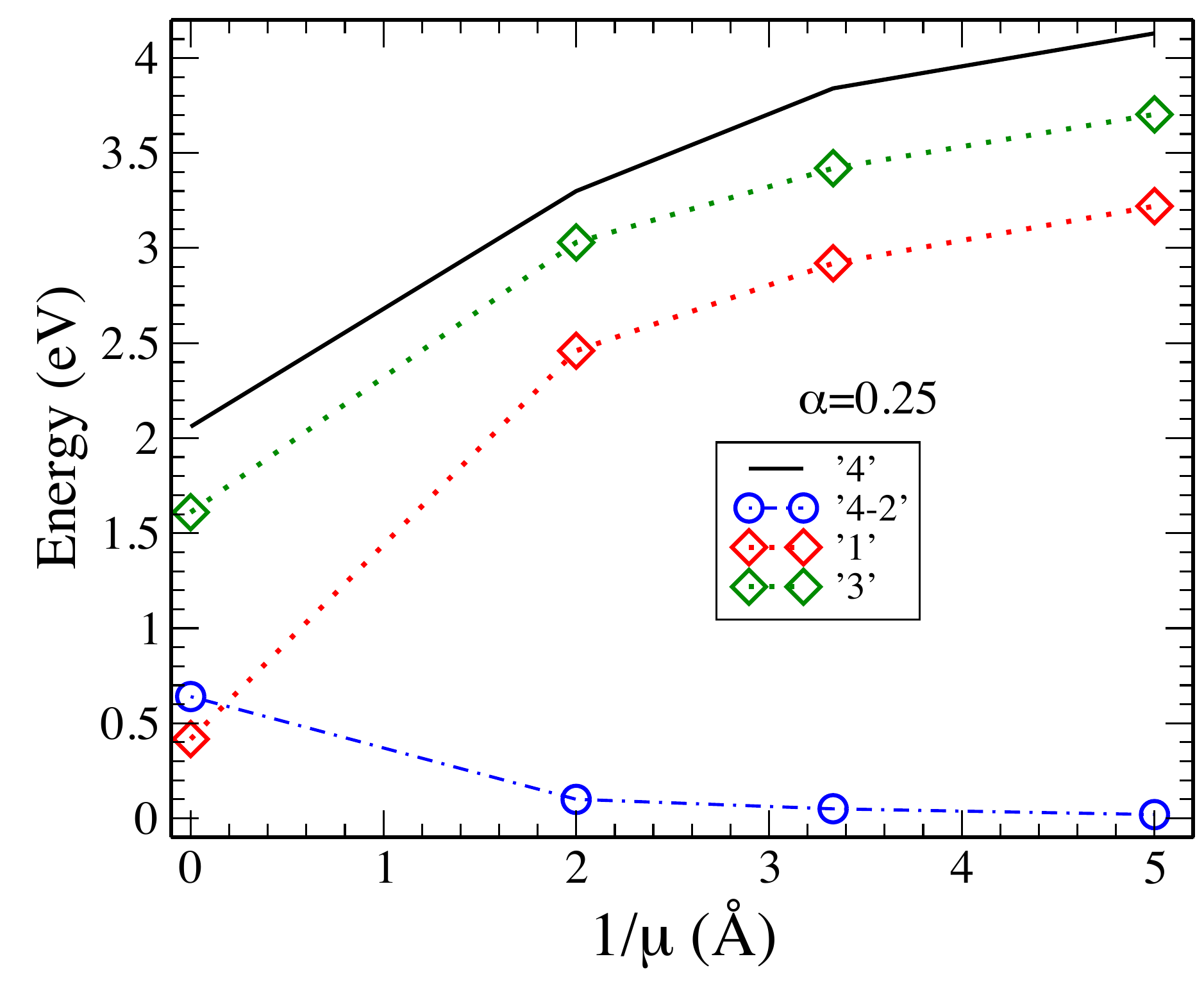}
\caption{(Color online) HSE transitions (see text, the scheme in Fig.\ref{fig1}) vs 1/$\mu$, with $\alpha$ fixed at 0.25}
\label{hsemu}
\end{figure}

In  this Section {we discuss the electronic structure of} AF CCTO as function of the relevant adjustable parameters of the various methods. {We consider the transitions defined in Fig.\ref{fig1}, namely} the fundamental gap, i.e. the `1' transition; the main charge transfer gap, i.e. the `4' transition; the upper valence-upper conduction gap, i.e. the `3' transition; and the minimum gap  between the  Cu-like lower conduction band and the Ti-like upper conduction band,  i.e. roughly the difference of the {`4' and `2' transitions, labeled `4--2'.  These} are  shown for GGA+U as function of the U--J parameter in  Fig.\ref{fig5},  for HSE as function of the   mixing parameter $\alpha$ in Fig.\ref{fig6} and of the screening parameter $\mu$ in  Fig.\ref{hsemu}. By construction,  plain GGA (which is quite similar to LDA in Fig.\ref{fig1}) is recovered in each of the limits of vanishing  U--J, $\alpha$, and 1/$\mu$. {The energies reported are edge-to-edge eigenvalue differences at the X point (for `1' and `3') and $\Gamma$ point (for `2' and `4').}

As  expected, all valence-to-conduction gaps increase rapidly with U--J, $\alpha$, and 1/$\mu$.  The `1' transition  increases fastest in all cases, and the conduction-conduction gap diminishes. This  means  that, relative to the valence band, the Cu-like lower conduction bands are pushed up in energy more than the Ti-like upper conduction bands,  opposite to what is observed in VPSIC.

This effect is especially strong  in GGA+U,  so much that the `4'--`2' transition becomes negative, i.e. the empty Cu band is pushed into the Ti band. 
This is because the U correction acts efficiently on the Cu-like bands by enhancing both spin polarization and orbital polarization. The former widens the `1' gap; the latter cleans up the  O $p$ valence from Cu-like character, pushing it down and widening somewhat the apparent charge transfer gaps.  We do not apply  a U on Ti. That  would  only have minor, if any,  effects: Ti orbitals are not spin-polarized, and U would only leverage orbital polarization, i.e. it would purge the valence and Cu-like states of what little Ti orbital content they have, thereby widening modestly, if at all, the relative gaps (at least for physically sensible values: of course one may always hope to get some effect for unphysically large U's). In conclusion, to obtain a `1' transition in the vicinity of 1 eV or less, U--J should be between 2 and 3, which is tiny by Cu-oxide standards.

In HSE, all gaps increase linearly with $\alpha$ (Fig.\ref{fig6}), with the `1' transition increasing faster than the others. 
As Fig.\ref{hsemu} shows, all the gaps decrease as $\mu$ increases, i.e. when  the Hartree-Fock exchange is gradually screened away.
At standard $\alpha$ the HSE functional gives a fairly good `4' transition, 4.1 eV at $\Gamma$. The `1' transition is  overestimated hugely at over 3 eV, whereas, depending on which experiments one believes and whatever uncertainties one may  attach to them, that gap is in the range of 0.5-1.5 eV at most. To obtain such a value one should either use a small $\alpha$ in the vicinity of 0.05, or, probably better, a more or less standard value like 0.15 to 0.25 with a larger than usual $\mu$ (as large as 1.6 \AA\, for $\alpha$=0.25). One way of stating this is that, compared to the upper bands, the Cu $d$ bands gets too large a ``correlation" correction (in the commonly-used, if questionable, sense of ``any correction needed beyond semilocal DFT'') from standard HSE, and that the correction should be more ``screened'' than it is,  consistently with the  smaller-than-usual  U--J mentioned above for GGA+U. An additional issue enhancing the sensitivity of Cu states to U-like corrections may have to do with the fact that the band structure of CCTO is dominated by long range hoppings.\cite{pick}  Whatever the final answer,  this overcorrection cannot be attributed (not straightforwardly, anyway) to self-interaction removal, which  operates --although in different guises-- in both HSE and VPSIC, and has moderate effects on the Cu states  in the latter.

It is worth reiterating that the  parameters in common use in the literature are those at the high end of the range considered here. U--J up to 8 or 9 eV is quite usual in Cu oxides,\cite{ybcochain} and $\alpha$=0.25, $\mu$=0.2 \AA$^{-1}$ is the standard  HSE recipe. For those values, both  GGA+U and HSE  produce an electronic structure whereby the Cu-like conduction states are way too high in energy, and the gap is too large by a factor of at least 3. {Used in their default setting}, GGA+U and  HSE  would predict  a fundamental gap `1' of 2.2 eV and 3.2 eV respectively, whereas experiments and VPSIC agree  that CCTO has a gap of less than 1 eV. Besides, the multiple absorptions involved in the CCTO spectra are not reproduced, as they end up being  squashed by the overcorrection into a single high-onset-energy transition.

\subsection{Discussion}

There are some  general conclusion to be drawn from the results just discussed. At the very bottom, GGA+U, HSE, and VPSIC are all semiempirical methods, in that they depend on some sort of parameter. One recognizes, though, that these parameters intervene very differently in each method. 

By construction, GGA+U is the most directly affected by its internal parameters. These can be estimated to some degree on a non-empirical basis from atomic quantities\cite{ggau} or from linear response,\cite{cococ} but in all cases they are  externally-determined system-dependent inputs  (occasionally even dependent on internal parameters or external conditions\cite{cococ2} within the same system), and not self-consistent and internal, so that in the end they are simply regarded as adjustable by most practitioneers. Whether this is admissible or desirable is as much a philosophical as an operational question that depends on the specific objectives of an investigation.  In the present case  we found that the consolidated parameter recipe simply does not seem to function. One may (we don't) elaborate further about which atom and shell  the correction should be applied to (e.g. would  U's on Ti or O $p$ make a difference: we argued above that they would not cure the problem), adding more parameters: this would probably bring us no nearer to a solution.
 
{As mentioned,} HSE's standard recipe has a theoretical foundation in the formulation of the functional, and has the merit of being system-independent (beside the practicality of including screening at the functional formulation level, and not a posteriori as in other hybrids\cite{pasq} based on, again,  empirical estimates\cite{fiobal} or models\cite{gb} of the screening).  That said,  the $\alpha$  and $\mu$ parameters  have been adjusted away from their standard value to cure various different issues (structure, electronic properties, magnetism, etc.) in many occasions, among which cupric oxide CuO.\cite{cuo-roc}

It is only fair to discuss in this context the parametric dependence built into VPSIC. For a detailed treatment we defer to the original work\cite{fs} and to a recent review,\cite{ff} which also discusses in detail the analogies and differences with GGA+U and hybrids. In short, the screening of atomic self-interaction corrections by the environment is described by a single parameter $\alpha_s$=1/2, a value based on a Slater transition-state argument.\cite{fs} The dependence
on  $\alpha_s$ of relevant quantities in solids  has  been studied,\cite{vpsicalpha} and the result is that $\alpha_s$=1/2 is indeed the optimal value on average over a vast class of materials. We systematically use that value,\cite{ff} hence effectively we do not  regard $\alpha_s$ as a parameter at all. In a case where a detailed comparison has been carried out,\cite{tmo} VPSIC has  been found to perform similarly to HSE;  discrepancies (and controversy) did occur in other cases, however, especially on cupric oxide.\cite{cuo-roc,cuo,cuo2} 

Since we are dealing with a titanate that is strongly characterized by cuprate-like electronic features,\cite{lcfo} it is  appropriate to recall the solid success record of VPSIC on the electronic and structural properties of Cu oxides of various composition and  dimensionality. It describes correctly the magnetic and insulating (anywhere from semiconducting to high-insulator) character  of  YBa$_2$Cu$_3$O$_{6}$,\cite{ybcochain}  monoclinic  CuO,\cite{cuo} GeCuO$_3$,\cite{gecuo} Ca-doped YCuO cuprate,\cite{caycuo} all of which are metals and non magnetic in GGA or LDA. This is further strong circumstantial evidence 
supporting the  use of  VPSIC  as reference for the other methods in CCTO, even if one were to  gloss over the experimental evidence discussed above.

\subsection{Quasiparticle corrections}
\label{gws}

It has long been customary to estimate quasiparticle energies as density-functional eigenvalues supplemented by ``self-energy corrections".\cite{gb,fiobal} In many materials, these corrections are dominated by  a ``scissor operator'', i.e. a $k$- and energy-independent relative shift of conduction and valence bands.  A simple empirically-determined scissor correction\cite{fiobal} is  $\Delta$$\simeq$9/$\varepsilon_{\infty}$ eV.  CCTO  has  a high-frequency dielectric constant $\varepsilon_{\infty}$=$\varepsilon_1$($\omega$=0)=12.6, so the  correction is $\Delta$$\sim$0.7 eV. The resulting total minimum gap is roughly 0.85 eV,   essentially in  the VPSIC (and experimental) ballpark. Thus, the VPSIC and empirical scissor give similar corrections to local functionals, despite being completely unrelated. 

Next we  calculate the same sort of correction   using  G$_{0}$W$_{0}$ non-selfconsistent quasiparticle energies.\cite{gw} The latter calculation is rather difficult to converge in general, and particularly for this large system. We use a softer O potential enabling a cutoff of 280 eV, which does not seem to affect the eigenvalues much. {(The use of the specialized potential PAW sets provided with VASP for GW calculations is prevented by their large energy cutoff; this should not be a serious problem, as the standard PAWs we use do  contain high-energy projectors and should perform rather well in the low-energy range we deal with here.)}  To assess convergence  in k and in the number of bands for the virtual-transitions summation we used  2$\times$2$\times$2  and 4$\times$4$\times$4 k-grids  and between 256 and 4092 bands. For a typical bulk material the latter choice would be overkill,  but our system has of order 130 occupied bands per spin channel, so this setting {seems necessary. Note also that that energy convergence in GW is more critical than in the standard joint-DOS calculations in the previous Sections:  in the latter, unoccupied bands are only used in the Kronig-Kramers relation, whereas in GW  they enter the evaluation of all energy-dependent parts of the self-energy.} 

\begin{center}
\begin{table}[ht]
\caption{Corrections to (semi)local transition energies (in eV, rounded to tenths of eV, labeled as in Fig.\ref{fig1}).}
\begin{tabular}{lcccc}
\hline
\hline
Transition & `1'	& `2' & `3'& `4'\\
\hline
$\Delta$(VPSIC--LDA) &0.5	& 0.7 & 1.1 & 1.4\\
$\Delta$(G$_0$W$_0$--GGA) &0.7	& 1.0 & 1.1 & 1.4\\
\hline
$\Delta$(empirical) &\multicolumn{4}{c}{0.7}\\
\hline\hline
\end{tabular}
\label{tab}
 \end{table}
\end{center}

{In the area of optics, at the simplest level of approximation, it has long been customary to address "beyond-LDA" (or GGA) corrections to eigenvalues or gap energies. We adopt this view and obtain the corrections  as the differences of eigenvalues within GW and GGA, and VPSIC and LDA respectively. This should keep bias at a minimum in the comparison  of the different technical settings  (potentials, chosen volume, DFT functionals, ...) and codes, besides hopefully providing some error cancellation.}

{The corrections are reported in Table \ref{tab} for the X point. From our partial convergence study vs bands and k-points, we judge that they  are converged to within 0.1 eV.} Interestingly,  the GW and empirical corrections agree well for the low energy gap; the GW and VPSIC corrections are also in decent overall agreement, and in particular they appear to depend on energy, i.e. higher bands are corrected more  than the lower ones. In particular, the differences between Cu and O $p$ upper valence, and  Cu  and Ti conduction states  are  the same in VPSIC and GW, i.e. the Cu $d$ empty band remains well clear of the Ti  {empty bands} in both cases. Also,  the lowest-energy GW gaps have the same character and order as in  VPSIC, the first gap being indirect between R and X and the second direct at X and less than 0.1 eV larger, i.e. the low-energy band topology appears similar in the two cases. Thus, overall, the corrections to local-functional eigenvalues provided by VPSIC are close to those of non-self-consistent one-shot GW, suggesting that much of the "beyond-local" correlation is indeed provided by VPSIC with  similar accuracy, at least in this material.\\ 

\section{Summary}

We examined the electronic structure of CaCu$_{3}$Ti$_{4}$O$_{12}$ as obtained via several different density-functional based methods, 
and proposed a  new interpretation of experiments  to the effect that four distinct transitions    contribute to the spectrum. The comparison of results from VPSIC calculations   with experiment is satisfactory, especially after we account for the effects of spin disorder, which does not close the fundamental gap but suppresses the intensity of the fundamental transition.  
 GGA+U and HSE at the standard values of  their internal parameters overestimate drastically the fundamental gap, hence the conclusion  that their corrections to the position of  the flat Cu $d$ bands should be more ``screened'' than they are. 
 On the other hand, the corrections to local- or semilocal- functional eigenvalues provided by VPSIC are close to those of non-self-consistent one-shot GW, suggesting that the "beyond-local" correlation is described by VPSIC with  similar accuracy.

\section*{Acknowledgments}

We thank P. Lunkenheimer for discussions, and especially for providing, and permitting the use of, the unpublished Tauc-relation data reported in Fig.\ref{lunkprivate}. Work supported in part  by MIUR PRIN 2010 {\it Oxide} and by Fondazione Banco di Sardegna and CINECA grants.


\begin{thebibliography}{99}
\bibitem{nota}
The `Mott' label is used here in the restricted sense of an odd-electron-count insulators (typically undergoing magnetic order-disorder transitions) to describe which one must describe reasonably   electron interactions across  energy scales from direct interaction, of order 1-10 eV, to exchange (Hund, super, double) of order 0.1-1 eV, and structural distortions, of order 0.01-0.1 eV. Within first-principles approaches,  only many-body perturbation-theory, hybrid--density-functional, and self-interaction-corrected density-functional methods seem able to reach this level of accuracy (of course, dynamical mean field theory does also, although it generally uses empirical inputs in the self-energy and, especially, about structure); all of these methods remove self interaction from  local density functionals--hence our interest in them in the present context.
\end{thebibliography}
\end{document}